\documentclass[letterpaper,twocolumn,american,showpacs,pre,aps,superscriptaddress]{revtex4}
\usepackage[latin1]{inputenc}
\usepackage{bm}
\usepackage{multirow,amssymb,amsbsy,amsmath}
\usepackage{stmaryrd}
\usepackage{graphicx}
\usepackage{hyperref}
\makeatletter
\usepackage{pifont}
\makeatother

\begin{document}

\title{Non-Markovian generalization of the Lindblad theory of open quantum systems}

\author{Heinz-Peter Breuer}

\email{breuer@physik.uni-freiburg.de}

\affiliation{Physikalisches Institut, Universit\"at Freiburg,
             Hermann-Herder-Strasse 3, D-79104 Freiburg, Germany}

\date{\today}

\begin{abstract}
A systematic approach to the non-Markovian quantum dynamics of
open systems is given by the projection operator techniques of
nonequilibrium statistical mechanics. Combining these methods with
concepts from quantum information theory and from the theory of
positive maps, we derive a class of correlated projection
superoperators that take into account in an efficient way
statistical correlations between the open system and its
environment. The result is used to develop a generalization of the
Lindblad theory to the regime of highly non-Markovian quantum
processes in structured environments.
\end{abstract}

\pacs{03.65.Yz, 05.70.Ln, 42.50.Lc, 03.65.Ta}

\maketitle

\section{Introduction}

The theoretical description of relaxation and decoherence
processes in open quantum systems often leads to a non-Markovian
dynamics which is determined by pronounced memory effects
\cite{TheWork}. Strong system-environment couplings
\cite{REIBOLD,HU}, correlations and entanglement in the initial
state \cite{INGOLD,BUZEK}, interactions with environments at low
temperatures and with spin baths \cite{LOSS}, finite reservoirs
\cite{GEMMER2005,GMM}, and transport processes in nano-structures
\cite{FOURIER} can lead to long memory times and to a failure of
the Markovian approximation.

A systematic approach to non-Markovian dynamics is provided by the
projection operator techniques \cite{NAKAJIMA,ZWANZIG,HAAKE} which
are extensively used in nonequilibrium thermodynamics and
statistical mechanics \cite{KUBO}. These techniques are based on
the introduction of a certain projection superoperator
${\mathcal{P}}$ which acts on the states of the total system. The
superoperator ${\mathcal{P}}$ is the mathematical expression for
the idea of the elimination of degrees of freedom from the
complete description of the states of the total system: If $\rho$
is the full density matrix of the composite system, the projection
${\mathcal{P}}\rho$ represents a certain approximation of $\rho$
which leads to a simplified effective description of the dynamics
through a reduced set of variables. The projection
${\mathcal{P}}\rho$ is therefore referred to as the relevant part
of the density matrix.

With the help of the projection operator techniques one derives
closed dynamic equations for the relevant part of the density
matrix. The equation of motion for ${\mathcal{P}}\rho$ can either
be the Nakajima-Zwanzig equation \cite{NAKAJIMA,ZWANZIG}, an
integrodifferential equation with a retarded memory kernel, or
else the time-convolutionless master equation, which is a
time-local differential equation of first order involving a
time-dependent generator \cite{SHIBATA}. In most cases these
equations are used as starting point for the derivation of
effective master equations through a systematic perturbation
expansion with respect to the strength of the system-environment
coupling.

In the standard approach to the dynamics of open systems one
chooses a projection superoperator which is defined by the
expression ${\mathcal{P}}\rho = \rho_S \otimes \rho_0$, where
$\rho_S={\mathrm{tr}}_E\rho$ represents the reduced density matrix
of the open system, ${\mathrm{tr}}_E$ denoting the trace over the
environmental Hilbert space, and $\rho_0$ is some fixed
environmental state. A superoperator of this form projects the
total state $\rho$ onto a tensor product state, i.~e., onto a
state without any statistical correlations between system and
environment. We emphasize that this ansatz does {\textit{not}}
imply (as is sometimes claimed) that one completely ignores all
system-environment correlations. It only presupposes that all
correlations which are present in the initial state or are
generated during the time-evolution can be treated as
perturbations within the framework of the projection operator
techniques.

The projection onto a tensor product state is widely used in
studies of open quantum systems. It is often applicable in the
case of weak system-environment couplings. Usually, the
perturbation expansion is restricted to the second order (known as
Born approximation), from which one derives, with the help of
certain further assumptions, a Markovian quantum master equations
in Lindblad form \cite{GORINI,LINDBLAD,SPOHN}. In this paper the
quantum dynamics of an open system is said to be non-Markovian if
the time-evolution of its reduced density matrix cannot be
described (to the desired degree of accuracy) by means of a closed
master equation with a time-independent generator in Lindblad
form.

A possible approach to large deviations from Markovian behavior
consists in carrying out the perturbation expansion to higher
orders in the system-environment coupling (several examples are
discussed in Ref.~\cite{TheWork}). However, this approach is often
limited by the increasing complexity of the resulting equations of
motion. Moreover, the perturbation expansion may not converge
uniformly in time, such that higher orders only improve the
quality of the approximation of the short-time behavior, but
completely fail in the long-time limit \cite{BBP}.

There is however a further promising strategy: To treat highly
non-Markovian processes in a more efficient way one can replace
the tensor product state used in the standard Born approximation
by a certain correlated system-environment state. This approach
has been proposed by Esposito and Gaspard \cite{GASPARD1,GASPARD2}
and by Budini \cite{BUDINI05} to derive effective master equations
within second order perturbation theory that describe strong
non-Markovian effects. It has been demonstrated in Ref.~\cite{BGM}
that this idea can be formulated in terms of a positive projection
superoperator ${\mathcal{P}}$ which projects any state onto a
correlated system-environment state, i.~e., onto a state that
contains statistical correlations between certain system and
environment states. This formulation allows an immediate
application of the projection operator techniques to correlated
system-environment states, and to carry out the perturbation
expansion to higher orders in a systematic way. An example is
discussed in Ref.~\cite{BGM}, where the master equations of second
and of fourth order corresponding to a correlated projection
superoperator have been constructed.

The application of a correlated projection superoperator implies
that the relevant part ${\mathcal{P}}\rho$ can no longer be
expressed in terms of the reduced density matrix alone. Hence,
employing a correlated projection superoperator one enlarges the
set of relevant variables to capture those statistical
correlations that are responsible for strong non-Markovian
effects.

In the present paper we discuss this idea of using correlated
projection superoperators in the analysis of non-Markovian
dynamics. On the basis of certain general physical conditions we
derive in Sec.~\ref{Sec-CPS} a representation theorem for a class
of correlated projection superoperators that are appropriate for
the application of the projection operator techniques.

A central problem of the theory of non-Markovian processes is the
formulation of appropriate master equations that preserve the
normalization and the positivity of the density matrix (see,
e.~g., the discussion in
Refs.~\cite{BARNETT,BUDINI04,LIDAR,MANISCALCO}). In
Sec.~\ref{DYNAMICS} we develop the general structure of such
master equations which results form the application of a
correlated projection superoperator. Given a superoperator that
projects onto a separable quantum state one can construct an
embedding of the underlying dynamics into a Lindblad dynamics on a
suitably extended state space. Employing this embedding we derive
a general class of physically acceptable master equations which
represents a generalization of the Lindblad theory to the regime
of highly non-Markovian quantum dynamics. Section \ref{CONCLU}
contains some conclusions.

\section{Correlated projection superoperators}\label{Sec-CPS}

\subsection{General conditions}\label{GEN-COND}

The Hilbert spaces of the open system $S$ and of its environment
$E$ are denoted by ${\mathcal{H}}_S$ and ${\mathcal{H}}_E$,
respectively. The state space of the composite system is given by
the tensor product
${\mathcal{H}}={\mathcal{H}}_S\otimes{\mathcal{H}}_E$. States of
the composite system are represented by density matrices $\rho$ on
${\mathcal{H}}$ satisfying $\rho \geq 0$ and ${\mathrm{tr}}\rho =
1$, where ${\mathrm{tr}}$ is the trace taken over the total state
space. The reduced density matrix $\rho_S$ of subsystem $S$ is
given by the partial trace taken over the Hilbert space
${\mathcal{H}}_E$, i.~e. $\rho_S={\mathrm{tr}}_E\rho$.
Correspondingly, the partial trace over ${\mathcal{H}}_S$ will be
denoted by ${\mathrm{tr}}_S$.

A superoperator ${\mathcal{P}}$ is a linear map ${\mathcal{O}}
\mapsto {\mathcal{P}} {\mathcal{O}}$ which takes any operator
${\mathcal{O}}$ on the total state space ${\mathcal{H}}$ to an
operator ${\mathcal{P}}{\mathcal{O}}$ on ${\mathcal{H}}$. We
consider here superoperators with the following properties.

{\textbf{1.}} The map ${\mathcal{P}}$ is a projection
superoperator:
\begin{equation} \label{PROJECTION}
 {\mathcal{P}}^2 = {\mathcal{P}}.
\end{equation}
It is this formal property that allows the application of the
projection operator techniques. For an efficient performance of
these techniques the projection ${\mathcal{P}}\rho$ should
represent a suitable approximation of $\rho$. A natural minimal
requirement is therefore that for any physical state $\rho$ the
projection ${\mathcal{P}}\rho$ is again a physical state, i.~e., a
positive operator with unit trace. This means that ${\mathcal{P}}$
is a positive and trace preserving map, namely $\rho \geq 0$
implies ${\mathcal{P}}\rho \geq 0$, and
${\mathrm{tr}}\{{\mathcal{P}}\rho\}={\mathrm{tr}}\rho$.

{\textbf{2.}} We consider projection superoperators of the
following general form:
\begin{equation} \label{ITIMESLAMBDA}
 {\mathcal{P}} = I_S \otimes \Lambda,
\end{equation}
where $I_S$ denotes the unit map acting on operators on
${\mathcal{H}}_S$, and $\Lambda$ is a linear map that takes
operators on ${\mathcal{H}}_E$ to operators on ${\mathcal{H}}_E$.
A projection superoperator of this form leaves the system $S$
unchanged and acts nontrivially only on the variables of the
environment $E$. As a consequence of the positivity of
${\mathcal{P}}$ and of condition (\ref{ITIMESLAMBDA}) the map
$\Lambda$ must be $N_S$-positive, where $N_S$ is the dimension of
${\mathcal{H}}_S$ (see, e.~g., Ref.~\cite{KOSSAKOWSKI}). In the
following we use the stronger condition that $\Lambda$ is
completely positive, because completely positive maps allow for a
simple mathematical characterization (see
Sec.~\ref{SEC-REPRESENTATION}).

We discuss the implications of these conditions. From
Eqs.~(\ref{PROJECTION}) and (\ref{ITIMESLAMBDA}) we get that
$\Lambda$ itself must be a projection, namely $\Lambda^2=\Lambda$.
Moreover, since ${\mathcal{P}}$ is trace-preserving, the map
$\Lambda$ must also be trace-preserving. Hence, we find that
$\Lambda$ represents a completely positive and trace-preserving
map (CPT map, or quantum channel) which operates on the variables
of the environment and has the property of a projection. A further
physically reasonable consequence of the positivity of $\Lambda$
and of Eq.~(\ref{ITIMESLAMBDA}) is that ${\mathcal{P}}$ maps
separable (classically correlated) states to separable states,
which means that the projection does not create entanglement
between system and environment.

An important goal is, of course, the determination of the reduced
density matrix $\rho_S$ of the open quantum system. Using
Eq.~(\ref{ITIMESLAMBDA}) and that $\Lambda$ is trace-preserving we
get the relation
\begin{equation} \label{CONSISTENCY}
 \rho_S \equiv {\mathrm{tr}}_E \rho = {\mathrm{tr}}_E
 \{ {\mathcal{P}} \rho \}.
\end{equation}
This relation connects the density matrix of the reduced system
with the projection of a given state $\rho$ of the total system.
It states that, in order to determine $\rho_S$, we do not really
need the full density matrix $\rho$, but only its projection
${\mathcal{P}}\rho$. Thus, ${\mathcal{P}}\rho$ contains the full
information needed to reconstruct the reduced system's state.

\subsection{Representation theorem}\label{SEC-REPRESENTATION}

We derive a representation theorem for the projection
superoperator ${\mathcal{P}}$ from the basic conditions formulated
in Sec.~\ref{GEN-COND} [see Eq.~(\ref{PROJECTION-GENFORM}) below].
Since $\Lambda$ is supposed to be a CPT map one could use, of
course, the Kraus-Stinespring representation
\cite{KRAUS,STINESPRING} for completely positive maps. However,
for our purposes another representation is much more appropriate,
which will be derived now.

We will use the following fact from linear algebra. Consider a
linear operator $L:V\mapsto V$ which acts on some Hilbert space
$V$ and has the property $L^2=L$. Then there exist linear
independent vectors $|f_i\rangle$ and linear independent vectors
$|e_i\rangle$ such that $\langle f_i | e_j\rangle=\delta_{ij}$ and
\begin{equation} \label{DEF-L}
 L|v\rangle = \sum_i |f_i\rangle\langle e_i|v\rangle
\end{equation}
for all $|v\rangle \in V$. Conversely, given two linear
independent sets $\{|f_i\rangle\}$ and $\{|e_i\rangle\}$ of
vectors in $V$ with $\langle f_i | e_j\rangle=\delta_{ij}$, then
Eq.~(\ref{DEF-L}) defines a linear operator with the property
$L^2=L$. Note that we neither require that the $|f_i\rangle$ or
the $|e_i\rangle$ are orthogonal, nor that $L$ is Hermitian.

Let us apply this fact to linear maps on the Hilbert-Schmidt
space, i.~e., we take $V$ to be the vector space of operators on
${\mathcal{H}}_E$ with the scalar product:
\[
 (X,Y) \equiv {\mathrm{tr}}_E\{ X^{\dagger} Y \}.
\]
Then we find that any linear map $\Lambda$ can be represented in
the form
\[
 \Lambda X = \sum_i B_i (A_i,X) =
 \sum_i {\mathrm{tr}}_E \{ A^{\dagger}_i X \} B_i
\]
with two sets $\{A_i\}$ and $\{B_i\}$ of linear independent
operators on ${\mathcal{H}}_E$, and that the condition
$\Lambda^2=\Lambda$ is satisfied if and only if $(B_i, A_j) =
\delta_{ij}$. Since $\Lambda$ preserves the Hermiticity of
operators, the operators $A_i$ and $B_i$ can be chosen to be
Hermitian. Hence, we obtain the representation:
\begin{equation} \label{LAMBDA}
 \Lambda X = \sum_i {\mathrm{tr}}_E \{ A_i X \} B_i,
\end{equation}
where $\{A_i\}$ and $\{B_i\}$ are two sets of linear independent
Hermitian operators satisfying:
\begin{equation} \label{BjAi}
 {\mathrm{tr}}_E \{ B_i A_j \} = \delta_{ij}.
\end{equation}
The condition that $\Lambda$ is trace-preserving takes the form:
\begin{equation} \label{TRACE-PRESERVING}
 \sum_i ({\mathrm{tr}}_E B_i) A_i  = I_E.
\end{equation}

Finally, we have to formulate the condition of the complete
positivity of the map $\Lambda$. A given map $\Lambda$ is
completely positive if and only if
\begin{equation} \label{COND-CP}
 (I_E \otimes \Lambda)(|\psi\rangle\langle\psi|) \geq 0,
\end{equation}
where
\[
 |\psi\rangle = \sum_{\alpha}
 |\alpha\rangle \otimes |\alpha\rangle
\]
is a maximally entangled vector in
${\mathcal{H}}_E\otimes{\mathcal{H}}_E$, and $\{|\alpha\rangle\}$
is an orthonormal basis for ${\mathcal{H}}_E$. To evaluate
condition (\ref{COND-CP}) we first note that
\begin{equation} \label{DERIV-1}
 (I_E\otimes\Lambda)(|\psi\rangle\langle\psi|) =
 \sum_{\alpha\beta} |\alpha\rangle\langle\beta|
 \otimes
 \Lambda(|\alpha\rangle\langle\beta|).
\end{equation}
On using the representation (\ref{LAMBDA}) one finds
\begin{equation} \label{DERIV-2}
 \Lambda(|\alpha\rangle\langle\beta|)
 = \sum_i \langle \beta | A_i | \alpha \rangle B_i
 = \sum_i \langle \alpha | A_i^T | \beta \rangle B_i,
\end{equation}
where $A_i^T$ denotes the transpose of the operator $A_i$ with
respect to the given basis $\{|\alpha\rangle\}$. Inserting
Eq.~(\ref{DERIV-2}) into Eq.~(\ref{DERIV-1}) we obtain
\begin{eqnarray*}
 (I_E\otimes\Lambda)(|\psi\rangle\langle\psi|) &=&
 \sum_i \sum_{\alpha\beta}
 |\alpha\rangle \langle \alpha | A_i^T | \beta \rangle
 \langle \beta | \otimes B_i \\
 &=& \sum_i A_i^T \otimes B_i.
\end{eqnarray*}
We conclude that a necessary and sufficient condition for
$\Lambda$ to be completely positive is given by the inequality
\begin{equation} \label{COND-POS}
 \sum_i A_i^T \otimes B_i \geq 0.
\end{equation}

Employing Eqs.~(\ref{LAMBDA}) and (\ref{ITIMESLAMBDA}) we obtain
the following representation for the projection superoperator
${\mathcal{P}}$,
\begin{equation} \label{PROJECTION-GENFORM}
 {\mathcal{P}}\rho = \sum_i {\mathrm{tr}}_E \{ A_i \rho \}
 \otimes B_i.
\end{equation}
Given observables $A_i$ and $B_i$ that satisfy Eqs.~(\ref{BjAi}),
(\ref{TRACE-PRESERVING}), and (\ref{COND-POS}), this equation
defines a projection superoperator which fulfills the general
conditions formulated in Sec.~\ref{GEN-COND}. Conversely, given a
projection which fulfills the conditions of Sec.~\ref{GEN-COND},
there exist observables $A_i$ and $B_i$ satisfying
Eqs.~(\ref{BjAi}), (\ref{TRACE-PRESERVING}), and (\ref{COND-POS})
such that Eq.~(\ref{PROJECTION-GENFORM}) holds. There are in
general many different sets of operators $A_i$, $B_i$ that
represent a given ${\mathcal{P}}$. If we have a particular set of
such operators, then the operators
\[
 A'_i = \sum_j u_{ij} A_j, \qquad
 B'_i = \sum_j v_{ij} B_j,
\]
represent the same projection, where $u=(u_{ij})$ and $v=(v_{ij})$
are real, non-singular matrices related by $u^Tv=I$.

\subsection{Examples}

Within the standard approaches one considers a projection
superoperator of the form
\begin{equation} \label{P-STANDARD}
 {\mathcal{P}}\rho = ({\mathrm{tr}}_E \rho ) \otimes \rho_0,
\end{equation}
where $\rho_0$ is any fixed environmental density matrix. Using a
projection of this form one assumes that the states of the total
system may be approximated by certain tensor products, describing
states without statistical dependencies between system and
environment. The projection (\ref{P-STANDARD}) naturally fits into
the general scheme developed above if we take a single $A=I_E$ and
a single $B=\rho_0$. The conditions (\ref{BjAi}),
(\ref{TRACE-PRESERVING}), and (\ref{COND-POS}) are then satisfied
and Eq.~(\ref{PROJECTION-GENFORM}) obviously reduces to
Eq.~(\ref{P-STANDARD})

In the general case, a projection ${\mathcal{P}}\rho$ of the form
of Eq.~(\ref{PROJECTION-GENFORM}) does not represent a simple
product state. We therefore call such ${\mathcal{P}}$ correlated
projection superoperators. They project onto states that contain
statistical correlations between the system $S$ and its
environment $E$. In the following we will consider the case that
one can find a representation of the projection with positive
operators $A_i \geq 0$ and $B_i \geq 0$. Equation (\ref{COND-POS})
is then trivially satisfied. Without restriction we may assume
that the $B_i$ are normalized to unit trace, such that condition
(\ref{TRACE-PRESERVING}) reduces to the simple form $\sum_i A_i =
I_E$. Under these conditions ${\mathcal{P}}$ projects any state
$\rho$ onto a state which can be written as a sum of tensor
products of positive operators. In the theory of entanglement
(see, e.~g., Ref.~\cite{ALBER}) such states are called separable
or classically correlated \cite{WERNER}. Using a projection
superoperator of this form, one thus tries to approximate the
total system's states by a classically correlated state. The
general representation of Eq.~(\ref{PROJECTION-GENFORM}) includes
the case of projection superoperators that project onto
inseparable, entangled quantum states. We will not pursue here
this possibility further, and restrict ourselves to positive $A_i$
and $B_i$ in the following.

A straightforward example for a separable projection superoperator
is obtained through the following construction
\cite{GASPARD1,BUDINI05,BGM}. We take any orthogonal decomposition
of the unit operator $I_E$ on the state space of the environment,
i.~e., a collection of ordinary projection operators $\Pi_i$ on
${\mathcal{H}}_E$ which satisfy
\[
 \Pi_i \Pi_j = \delta_{ij} \Pi_j, \qquad \sum_i \Pi_i = I_E.
\]
Then we choose:
\[
 A_i = \Pi_i, \qquad
 B_i = \frac{\Pi_i\rho_0\Pi_i}{{\mathrm{tr}}_E\{\Pi_i\rho_0\}}.
\]
It is easy to verify that with this choice the conditions
(\ref{BjAi}), (\ref{TRACE-PRESERVING}), and (\ref{COND-POS}) are
satisfied. The explicit form of the projection superoperator is
given by
\begin{equation} \label{PROJ-NEW}
 {\mathcal{P}} \rho = \sum_i
 {\mathrm{tr}}_E \left\{ \Pi_i \rho \right\} \otimes
 \frac{\Pi_i\rho_0\Pi_i}{{\mathrm{tr}}_E\{\Pi_i\rho_0\}}.
\end{equation}

\subsection{Relevant states and observables}\label{REL-OBS}

Given a projection superoperator we define the relevant states as
the states in the range of ${\mathcal{P}}$, i.~e., for which the
relation
\[
 {\mathcal{P}}\rho_{\mathrm{rel}} = \rho_{\mathrm{rel}}
\]
holds. We see that these states are of the form
\[
 \rho_{\mathrm{rel}} = \sum_i \rho_i \otimes B_i,
\]
where the $\rho_i$ may be any positive matrices such that $\sum_i
{\mathrm{tr}}_S\rho_i=1$. Hence, the manifold of the relevant
states is determined by the operators $B_i$.

One can transfer this concept from states (Schr\"odinger picture)
to observables (Heisenberg picture). A Hermitian operator
${\mathcal{O}}_{{\mathrm{rel}}}$ on the total state space is said
to be a relevant observable if the relation
\[
 {\mathrm{tr}}\{ {\mathcal{O}}_{{\mathrm{rel}}} ({\mathcal{P}}\rho) \}
 = {\mathrm{tr}} \{ {\mathcal{O}}_{{\mathrm{rel}}} \rho \}
\]
holds true for all states $\rho$. This means that the expectation
value of a relevant observable in any state of the composite
system is left unchanged under the application of the projection
superoperator ${\mathcal{P}}$.

Introducing the adjoint map ${\mathcal{P}}^{\dagger}$ (defined
with respect to the Hilbert-Schmidt scalar product for operators
on the total state space) we can reformulate this condition as
\[
 {\mathcal{P}}^{\dagger} {\mathcal{O}}_{{\mathrm{rel}}}
 = {\mathcal{O}}_{{\mathrm{rel}}}.
\]
Hence, the relevant observables are those observables which are
invariant under the application of the adjoint projection. From
the representation (\ref{PROJECTION-GENFORM}) we get:
\begin{equation} \label{PROJECTION-GENFORM-ADJOINT}
 {\mathcal{P}}^{\dagger} {\mathcal{O}} = \sum_i {\mathrm{tr}}_E
 \{ B_i {\mathcal{O}} \} \otimes A_i,
\end{equation}
from which we infer that the relevant observables must be of the
form
\begin{equation} \label{O-REL}
 {\mathcal{O}}_{{\mathrm{rel}}}
 = \sum_i {\mathcal{O}}^i_S \otimes A_i,
\end{equation}
where the ${\mathcal{O}}^i_S$ are arbitrary observables of the
subsystem $S$. The structure of the relevant observables is thus
determined by the operators $A_i$.

\section{Dynamics}\label{DYNAMICS}

\subsection{General formulation}
The dynamics of the total system is given by
\begin{equation} \label{NEUMANN}
 \rho(t) = U(t)\rho(0)U^{\dagger}(t),
\end{equation}
where $U(t)$ denotes the unitary time-evolution operator. Given a
projection superoperator of the form of
Eq.~(\ref{PROJECTION-GENFORM}) one introduces the dynamical
variables:
\begin{equation} \label{RHO-I}
 \rho_i(t) = {\mathrm{tr}}_E \left\{ A_i \rho(t) \right\}.
\end{equation}
Since the $A_i$ are positive operators, we have $\rho_i(t)\geq 0$,
and by use of Eq.~(\ref{CONSISTENCY}) and  of the normalization
${\mathrm{tr}}_E B_i = 1$ we find that the reduced density matrix
is obtained from
\[
 \rho_S(t) = \sum_i \rho_i(t).
\]
The normalization condition reads:
\begin{equation} \label{NORMIERUNG}
 {\mathrm{tr}}_S \rho_S(t) = \sum_i {\mathrm{tr}}_S \rho_i(t) = 1.
\end{equation}
Hence, the state of the reduced system is determined by a certain
set of unnormalized density matrices $\rho_i(t)$.

We consider initial conditions of the following form,
\begin{equation} \label{INIT}
 \rho(0) = {\mathcal{P}}\rho(0) = \sum_i \rho_i(0) \otimes B_i.
\end{equation}
This equation means that the initial state belongs to the manifold
of the relevant states. As a consequence there is no inhomogeniety
in the Nakajima-Zwanzig or the time-convolutionless master
equation, although $\rho(0)$ describes a correlated
system-environment state. On using Eqs.~(\ref{NEUMANN}),
(\ref{RHO-I}), and (\ref{INIT}) we get:
\begin{equation} \label{RHO-I-EVOL}
 \rho_i(t) = \sum_j {\mathrm{tr}}_E \left\{
 A_i U(t) \rho_j(0) \otimes B_j U^{\dagger}(t) \right\}.
\end{equation}

To be specific we assume that the index $i$ takes on the values
$i=1,2,\ldots,n$. We define a vector $\varrho$ whose components
are given by the dynamical variables $\rho_i$:
\begin{equation} \label{VEC-RHO}
 \varrho = (\rho_1,\rho_2,\ldots,\rho_n).
\end{equation}
Equation (\ref{RHO-I-EVOL}) then defines a dynamical
transformation of the form
\[
 V_t: \varrho(0) \mapsto \varrho(t),
\]
i.~e., we get a one-parameter family of dynamical maps $\{ V_t
\mid t \geq 0 \}$, where $V_0$ is equal to the identity map. For
each fixed $t$ the map $V_t$ transforms any initial vector
$\varrho(0)$ with positive components $\rho_i(0)\geq 0$ into some
vector $\varrho(t)$ with positive components $\rho_i(t)\geq 0$. Of
course, this transformation also preserves the normalization
condition (\ref{NORMIERUNG}).

It is important to emphasize that $V_t$ is {\textit{not}} a
quantum dynamical map in the conventional sense of the theory of
open systems, simply because it is {\textit{not}} an operation on
the space of states of the reduced system, but rather a map on the
space of vectors $\varrho$. The transition from $\varrho(0)$ to
the reduced density matrix $\rho_S(0)=\sum_i \rho_i(0)$ is
connected with a loss of information on the initial correlations,
such that from the mere knowledge of $\rho_S(0)$ the dynamical
behavior cannot be reconstructed, in general. In other words, the
evolution from $\rho_S(0)$ to $\rho_S(t)$ for $t>0$ is not a map,
i.~e., there is no prescription which assigns to each $\rho_S(0)$
a unique $\rho_S(t)$.

\subsection{Structure of non-Markovian master
            equations}\label{STRUCTURE}

The application of the time-convolutionless projection operator
technique leads to a closed dynamic equation for the relevant part
of the density matrix:
\[
 \frac{d}{dt} {\mathcal{P}}\rho(t) =
 {\mathcal{K}}^t({\mathcal{P}}\rho(t)),
\]
where ${\mathcal{K}}^t$ is a linear generator which depends, in
general, explicitly on time. If one uses a projection of the form
of Eq.~(\ref{PROJECTION-GENFORM}) this yields a system of
equations of motion for the densities $\rho_i(t)$,
\[
 \frac{d}{dt} \rho_i =
 {\mathcal{K}}^t_i(\rho_1,\ldots,\rho_n), \qquad
 i=1,2,\ldots,n,
\]
with an explicitly time-dependent generator ${\mathcal{K}}^t_i$
for each $i$ which depends linearly on the input arguments
$\rho_1,\ldots,\rho_n$. Let us suppose that these generators may
be approximated by time-independent generators such that we can
write
\begin{equation} \label{MASTER-TCL}
 \frac{d}{dt} \rho_i =
 {\mathcal{K}}_i(\rho_1,\ldots,\rho_n).
\end{equation}
The family $V_t$ then represents a semigroup of dynamical
transformations. Of course, this semigroup assumption for $V_t$
does not imply that the dynamics of the reduced density matrix
yields a semigroup.

A typical situation in which the semigroup assumption is valid is
given by a projection superoperator of the form of
Eq.~(\ref{PROJ-NEW}), where the $\Pi_i$ project onto certain
energy bands of the environment, describing a structured
reservoir. The semigroup assumption then presupposes that all
intra- and inter-band transitions may be described by means of
time-independent rates obtained from Fermi's golden rule
\cite{GASPARD1,BUDINI05,BGM}.

Our aim is to determine the general structure of the generators
${\mathcal{K}}_i$. To this end, we demand that
Eq.~(\ref{MASTER-TCL}) preserves the positivity of all components
$\rho_i$, i.~e., given positive initial data $\rho_i(0)\geq 0$ the
corresponding solution satisfies $\rho_i(t)\geq 0$ for all times.
A convenient way of formulating the dynamical transformation is
the following. We introduce an auxiliary Hilbert space ${\mathbb
C}^n$ and an orthonormal basis $\{ |i\rangle \}$ for this space.
Then we can identify the vector $\varrho$ introduced in
Eq.~(\ref{VEC-RHO}) with a density matrix on the extended space
${\mathcal{H}}_S\otimes{\mathbb C}^n$:
\begin{equation} \label{EXTENDED-SPACE}
 \varrho = \sum_i \rho_i \otimes |i\rangle\langle i|.
\end{equation}
This density matrix can be regarded as a block diagonal matrix
with blocks $\rho_i$ along the main diagonal. Moreover, the
reduced density matrix $\rho_S$ is obtained by the partial trace
of $\varrho$ taken over the auxiliary space. Hence, the auxiliary
space represents an additional degree of freedom which expresses
the statistical correlations introduced through the projection
superoperator ${\mathcal{P}}$.

The dynamical transformation $V_t$ can be viewed as a CPT
operation on the extended space that preserves the block diagonal
structure. To guarantee the conservation of positivity we
therefore impose the condition that there exists a Lindblad
generator ${\mathcal{L}}$ on the extended space which also
preserves the block diagonal structure, i.~e., which has the
property
\begin{equation} \label{L-PROP}
 {\mathcal{L}} \left( \sum_i \rho_i \otimes |i\rangle\langle i|
 \right) = \sum_i {\mathcal{K}}_i(\rho_1,\ldots,\rho_n)
 \otimes |i\rangle\langle i|.
\end{equation}
If such a Lindblad generator exists, the solution of
Eq.~(\ref{MASTER-TCL}) can be written as
\begin{equation} \label{EMBEDDING}
 \sum_i \rho_i(t) \otimes |i\rangle\langle i|
 = e^{{\mathcal{L}}t} \left(
 \sum_i \rho_i(0) \otimes |i\rangle\langle i| \right).
\end{equation}
This implies the required conservation of positivity of the
components $\rho_i(t)$ for all times $t\geq 0$. In mathematical
terms, Eq.~(\ref{EMBEDDING}) can be interpreted as an embedding of
the non-Markovian dynamics into a Lindblad dynamics on the
extended state space. We now demonstrate that the simple condition
expressed by Eq.~(\ref{L-PROP}) fixes the structure of the
generators ${\mathcal{K}}_i$ to a large extend.

{\textit{Theorem.}} A Lindblad generator ${\mathcal{L}}$ on the
extended state space with the property (\ref{L-PROP}) exists if
and only if the generators ${\mathcal{K}}_i$ are of the special
form
\begin{eqnarray} \label{GEN-K}
 \lefteqn{ {\mathcal{K}}_i(\rho_1,\ldots,\rho_n) } \\
 &=& -i \left[ H^i,\rho_i \right] + \sum_{j \lambda} \left(
 R^{ij}_{\lambda} \rho_j R^{ij\dagger}_{\lambda}
 - \frac{1}{2} \left\{ R^{ji\dagger}_{\lambda} R^{ji}_{\lambda},
 \rho_i\right\} \right) \nonumber
\end{eqnarray}
with Hermitian operators $H^i$ and arbitrary system operators
$R^{ij}_{\lambda}$.

{\textit{Proof.}} Assume that a Lindblad generator ${\mathcal{L}}$
on the extended state space with the property (\ref{L-PROP}) is
given. As for any Lindblad generator we have
 \begin{equation} \label{LINDBLAD-GEN}
 {\mathcal{L}}(\varrho) = -i \left[ H,\varrho \right]
 + \sum_{\lambda} \left( R_{\lambda} \varrho R^{\dagger}_{\lambda}
 - \frac{1}{2}
 \left\{ R^{\dagger}_{\lambda} R_{\lambda}, \varrho \right\}
 \right),
\end{equation}
were $H=H^{\dagger}$ and the $R_{\lambda}$ are operators on the
extended state space. These operators can always be written as
sums over tensor products:
\begin{eqnarray*}
 H &=& \sum_{ij} H^{ij} \otimes |i\rangle\langle j|, \\
 R_{\lambda} &=& \sum_{ij} R^{ij}_{\lambda} \otimes |i\rangle\langle j|.
\end{eqnarray*}
Substituting these relations into Eq.~(\ref{LINDBLAD-GEN}) and
using the expression (\ref{EXTENDED-SPACE}) for $\varrho$ one
easily shows that
\[
 {\mathcal{L}}(\varrho) =
 \sum_{ik} D^{ik} \otimes |i\rangle\langle k|,
\]
where
\begin{eqnarray} \label{D-IK}
 D^{ik} &=& -i \left( H^{ik}\rho_k - \rho_i H^{ik} \right) \\
 &+& \sum_{j\lambda} \left( R^{ij}_{\lambda} \rho_j R^{kj\dagger}_{\lambda}
 - \frac{1}{2} R^{ji\dagger}_{\lambda} R^{jk}_{\lambda} \rho_k
 - \frac{1}{2} \rho_i R^{ji\dagger}_{\lambda} R^{jk}_{\lambda}
 \right). \nonumber
\end{eqnarray}
Hence, in order for condition (\ref{L-PROP}) to be satisfied we
must have $D^{ik}=0$ for all $i\neq k$, and, in particular,
$D^{ii}={\mathcal{K}}_i$. Setting $i=k$ in Eq.~(\ref{D-IK}) we get
the form (\ref{GEN-K}) for the generator ${\mathcal{K}}_i$, where
$H^{ii}=H^i$ is Hermitian.

Suppose now that ${\mathcal{K}}_i$ is of the form of
Eq.~(\ref{GEN-K}). Then we define operators on the extended state
space through
\begin{eqnarray*}
 H &=& \sum_{i} H^i \otimes |i\rangle\langle i|, \\
 S^{ij}_{\lambda} &=& R^{ij}_{\lambda} \otimes |i\rangle\langle j|,
\end{eqnarray*}
where $H$ is Hermitian. With the help of these operators we define
a Lindblad generator by means of
\[
 {\mathcal{L}}(\varrho) = -i \left[ H,\varrho \right]
 + \sum_{ij\lambda} \left( S^{ij}_{\lambda} \varrho S^{ij\dagger}_{\lambda}
 - \frac{1}{2}
 \left\{ S^{ij\dagger}_{\lambda} S^{ij}_{\lambda}, \varrho \right\}
 \right).
\]
It is easy to check that this Lindblad generator indeed has the
required property (\ref{L-PROP}). This proves the theorem.

\subsection{Discussion}
Inserting Eq.~(\ref{GEN-K}) into Eq.~(\ref{MASTER-TCL}) we obtain
the master equation
\begin{eqnarray} \label{GEN-MASTER}
 \frac{d}{dt} \rho_i &=& -i \left[ H^i,\rho_i \right] \nonumber \\
 && + \sum_{j \lambda} \left(
 R^{ij}_{\lambda} \rho_j R^{ij\dagger}_{\lambda}
 - \frac{1}{2} \left\{
 R^{ji\dagger}_{\lambda} R^{ji}_{\lambda}, \rho_i\right\} \right).
\end{eqnarray}
Under the condition of the existence of an embedding into a
Lindblad dynamics on the extended state space [see
Eq.~(\ref{EMBEDDING})], this is the general form for the equations
of motion of the dynamical variables $\rho_i(t)$, where the $H^i$
are arbitrary Hermitian operators, and the $R^{ij}_{\lambda}$ are
arbitrary system operators.

\subsubsection{Examples}\label{EXAMPLES}
Several master equations proposed recently in the literature are
of the general form of Eq.~(\ref{GEN-MASTER}). The simplest
special case of this equation is obtained if we have only a single
component $\rho_1=\rho_S$ such that we can omit the indices $i$
and $j$. The master equation (\ref{GEN-MASTER}) then obviously
reduces to an ordinary master equation for the reduced density
matrix in Lindblad form. Thus, Eq.~(\ref{GEN-MASTER}) can be
viewed as a generalization of the
Gorini-Kossakowski-Sudarshan-Lindblad theorem
\cite{GORINI,LINDBLAD}.

If we choose $R^{ij}_{\lambda}=\delta_{ij}R^i_{\lambda}$ the
master equation (\ref{GEN-MASTER}) takes the form
\[
 \frac{d}{dt} \rho_i = {\mathcal{L}}_i(\rho_i).
\]
Hence, we get an uncoupled system of equations of motion with a
Lindblad generator ${\mathcal{L}}_i$ for each component $\rho_i$.
Although each component $\rho_i(t)$ follows its own Markovian-type
dynamics, the dynamics of the reduced density matrix $\rho_S(t)$
is generally highly non-Markovian. Master equations of this
uncoupled form have been derived by Budini and applied to various
models for the dynamics of open systems in structured reservoirs
\cite{BUDINI05}.

\begin{figure}[htb]
\includegraphics[width=\linewidth]{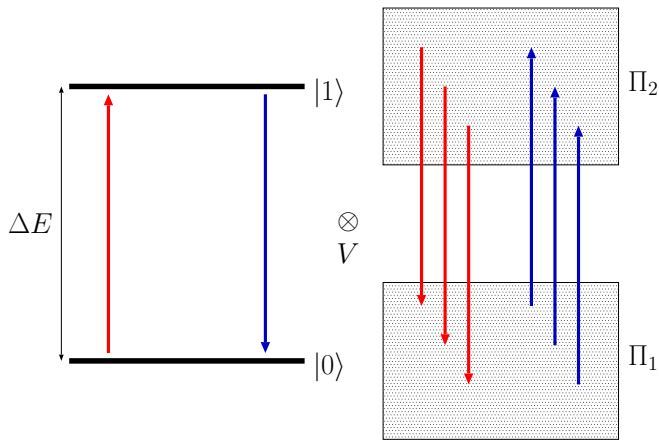}
\caption{A two-state system which is coupled to an environment
consisting of two energy bands.\label{model-fig}}
\end{figure}

In the general case an equation of the form (\ref{GEN-MASTER})
involves a coupling between the components $\rho_i$. An example of
such an equation has been derived in \cite{BGM} from a specific
microscopic system-environment model. The model describes a
two-state system with ground state $|0\rangle$, excited state
$|1\rangle$ and level distance $\Delta E$, which is coupled to an
environment \cite{GEMMER2005}. The environment consists of a large
number of energy levels arranged in two energy bands of width
$\delta\varepsilon$ (see Fig.~\ref{model-fig}). The lower energy
band contains $N_1$ levels, the upper band $N_2$ levels.

The total Hamiltonian of the model is taken to be
\begin{equation} \label{H-MODEL}
 H = H_S + H_E + V.
\end{equation}
$H_S=\Delta E \sigma_+\sigma_-$ is the free system Hamiltonian,
where $\sigma_\pm$ denote the usual raising and lowering operators
of the two-state system. The free Hamiltonian of the environment
is given by
\[
 H_E = \sum_{n_1} \frac{\delta\varepsilon}{N_1} n_1 |n_1\rangle\langle n_1|
 + \sum_{n_2}
 \left( \Delta E + \frac{\delta\varepsilon}{N_2}n_2 \right)
 |n_2\rangle\langle n_2|,
\]
and the system-environment interaction is described by
\[
 V = \lambda \sum_{n_1,n_2} c(n_1,n_2) \sigma_+
 |n_1\rangle\langle n_2| + {\mbox{h.c.}}
\]
The index $n_1$ labels the levels of the lower energy band and
$n_2$ the levels of the upper band. The overall strength of the
interaction is parameterized by the constant $\lambda$. The
coupling constants $c(n_1,n_2)$ are independent and identically
distributed complex Gaussian random variables with zero mean and
unit variance.

We employ a projection superoperator of the form \cite{BGM}
\begin{eqnarray} \label{NEWPROJECTION}
 {\mathcal{P}} \rho &=&
 {\mathrm{tr}}_E \left\{ \Pi_1 \rho \right\} \otimes \frac{1}{N_1} \Pi_1
 + {\mathrm{tr}}_E \left\{ \Pi_2 \rho \right\} \otimes \frac{1}{N_2}
 \Pi_2 \nonumber \\
 &\equiv& \rho_1 \otimes \frac{1}{N_1} \Pi_1
 + \rho_2 \otimes \frac{1}{N_2} \Pi_2,
\end{eqnarray}
where $\Pi_1$ ($\Pi_2$) denotes the projection onto the lower
(upper) environmental energy band. This is a projection of the
form given in Eq.~(\ref{PROJ-NEW}). It projects onto a state in
which the environmental state $\Pi_i/N_i$ is correlated with the
system state $\rho_i/{\mathrm{tr}}_S\rho_i$. The second order of
the time-convolutionless projection operator technique leads to
the master equation (written in the interaction picture)
\begin{eqnarray}
 \frac{d}{dt} \rho_1 &=&
 \gamma_1 \sigma_+ \rho_2 \sigma_-
 -\frac{\gamma_2}{2} \{ \sigma_+\sigma_-, \rho_1 \},
 \label{MASTER1} \\
 \frac{d}{dt} \rho_2 &=&
 \gamma_2 \sigma_- \rho_1 \sigma_+
 -\frac{\gamma_1}{2}\{ \sigma_-\sigma_+, \rho_2 \},
 \label{MASTER2}
\end{eqnarray}
where $\gamma_{1,2}=2\pi\lambda^2N_{1,2}/\delta\varepsilon$. This
is a coupled system of first-order differential equations for the
two unnormalized density matrices $\rho_1$ and $\rho_2$. It can be
written in the form of Eq.~(\ref{GEN-MASTER}) by taking $H^i=0$,
$R^{11}=R^{22}=0$, $R^{12}=\sqrt{\gamma_1}\sigma_+$, and
$R^{21}=\sqrt{\gamma_2}\sigma_-$. As has been demonstrated in
Ref.~\cite{BGM} through a comparison with numerical simulations of
the full Schr\"odinger equation, this master equation yields an
excellent approximation of the reduced system's dynamics.

The second term on the right-hand side of Eq.~(\ref{MASTER1})
describes changes of $\rho_1$ which are due to downward
transitions of the two-state system. These are only possible if
the lower band of the environment is populated, i.~e., if the
environment is in the state $\Pi_1/N_1$ which is correlated with
$\rho_1$. For this reason the second term of Eq.~(\ref{MASTER1})
involves the density $\rho_1$. Correspondingly, the first term on
the right-hand side of Eq.~(\ref{MASTER1}) describes changes of
$\rho_1$ caused by excitations of the two-state system. Such
excitations are only possible if the environment is in the state
$\Pi_2/N_2$ which is correlated with $\rho_2$. Therefore, it is
the density $\rho_2$ that enters the first term of
Eq.~(\ref{MASTER1}). Analogous statements hold for
Eq.~(\ref{MASTER2}). Hence, we see that the transitions described
by Eqs.~(\ref{MASTER1}) and (\ref{MASTER2}) exactly conserve the
total number of excitations (see Sec.~\ref{CONS}).

Let us investigate the behavior of the excited state population
$p_e(t)$ which is defined by the expression
\[
 p_e(t) = \langle 1 | \rho_1(t) + \rho_2(t) | 1 \rangle.
\]
Equations (\ref{MASTER1}) and (\ref{MASTER2}) lead to
\begin{eqnarray} \label{POP}
 \frac{d}{dt} p_e(t) &=& -(\gamma_1+\gamma_2) p_e(t)
 + \gamma_1 p_e(0) \nonumber \\
 &~&  + \gamma_2 \langle 1 | \rho_2(0) | 1 \rangle
      + \gamma_1 \langle 0 | \rho_2(0) | 0 \rangle.
\end{eqnarray}
This equation exhibits a strong non-Markovian character because
the initial data appear as inhomogeneous terms on its right-hand
side. These terms express a pronounced memory effect: They imply
that the dynamics never forgets its initial condition, i.~e., that
the process is governed by an infinite memory time. A further
remarkable feature of Eq.~(\ref{POP}) derives from the fact that
the initial data on its right-hand side cannot be expressed solely
in terms of the matrix elements of the reduced density matrix.
This means that at any time $t$, and even in the limit
$t\rightarrow\infty$, the process is strongly influenced by the
statistical correlations of the initial state. Hence, the
influence of the initial correlations never dies out and is
present even in the stationary state.

Recently, Esposito and Gaspard \cite{GASPARD1,GASPARD2} have
derived a master equation from a microscopic system-reservoir
model within second order perturbation theory, which is also of
the form of Eq.~(\ref{GEN-MASTER}). In their derivation the index
$i$ plays the role of the energy $\varepsilon$ of the reservoir.
The corresponding density matrix $\rho_i\equiv\rho_{\varepsilon}$
describes a system state which is correlated with a certain
reservoir state of energy $\varepsilon$. If the open system
represents again a two-state system, the master equation of
Esposito and Gaspard can be written as \footnote{See Sec. III A of
Ref.~\cite{GASPARD1}. The authors of this work use a continuous
energy variable $\varepsilon$. Here we keep the discrete notation
and employ a slightly different treatment of the Lamb shift
terms.}
\begin{eqnarray*}
 \frac{d}{dt} \rho_{\varepsilon} &=&
 -i \left[ H^{\varepsilon},\rho_{\varepsilon} \right] \\
 && \hspace{-8mm}
 + \sum_{\varepsilon'} \left(
 \gamma_1(\varepsilon,\varepsilon') \sigma_+ \rho_{\varepsilon'} \sigma_-
 - \frac{\gamma_2(\varepsilon',\varepsilon)}{2} \left\{
 \sigma_+\sigma_-, \rho_{\varepsilon} \right\} \right. \\
 && \left.
 + \gamma_2(\varepsilon,\varepsilon') \sigma_- \rho_{\varepsilon'} \sigma_+
 - \frac{\gamma_1(\varepsilon',\varepsilon)}{2} \left\{
 \sigma_-\sigma_+, \rho_{\varepsilon} \right\}
 \right).
\end{eqnarray*}
Here, $\gamma_{1,2}(\varepsilon,\varepsilon')$ are certain
transition rates which are determined by the parameters of the
microscopic model, and $H^{\varepsilon}$ is the system Hamiltonian
including an $\varepsilon$-dependent Lamb-type energy
renormalization. One easily checks that this master equation can
indeed be brought into the general form of Eq.~(\ref{GEN-MASTER})
by taking $R_1^{\varepsilon\varepsilon'}=
\sqrt{\gamma_1(\varepsilon,\varepsilon')}\sigma_+$ and
$R_2^{\varepsilon\varepsilon'}=
\sqrt{\gamma_2(\varepsilon,\varepsilon')}\sigma_-$.

A master equation of the general form (\ref{GEN-MASTER}) has been
derived recently by Budini \cite{BUDINI06}. This author suggests
introducing an extra degree of freedom $U$ which modulates the
interaction between the reduced system $S$ and its environment
$E$. Under the assumptions that $E$ may be treated as a Markovian
reservoir and that the dynamics of the populations decouples from
the dynamics of the coherences of $U$, one arrives at the master
equation (\ref{GEN-MASTER}). In a certain sense the introduction
of an additional degree of freedom $U$ corresponds to the extended
state space ${\mathcal{H}}_S\otimes{\mathbb C}^n$ used in
Sec.~\ref{STRUCTURE} to construct the embedding into a Lindblad
dynamics. An advantage of the present formulation is that it
avoids the use of a microscopic model for the extra degree of
freedom and that it clearly shows the basic physical condition
underlying the master equation (\ref{GEN-MASTER}): This condition
is the existence of an effective representation of the dynamics
through a projection onto separable, classically correlated
quantum states.

What are the physical conditions under which the use of a standard
projection onto a tensor product state is not sufficient to
correctly describe the reduced system's dynamics for a given
system-environment model? While this important question seems to
be difficult to answer in full generality and certainly deserves
further investigations, important hints can be obtained already by
an investigation of the time-convolutionless perturbation
expansion \cite{SHIBATA,TheWork}. If the two-point environmental
correlation functions do not decay rapidly in time the second
order of the expansion cannot, of course, be expected to give an
accurate description of the dynamics. For instance, this situation
arises for the spin star model discussed in Ref.~\cite{BBP}, where
the second-order generator of the master equation increases
linearly with time such that the Born-Markov approximation simply
does not exist. The example investigated in Ref.~\cite{BGM}
demonstrates that there are even cases in which the standard
Markov condition is satisfied although the product-state
projection completely fails if one truncates the expansion at any
finite order. In such cases strong non-Markovian dynamics is
induced through the behavior of higher-order correlation
functions. Hence, one can judge the quality of a given projection
superoperator only by an investigation of the structure of higher
orders of the expansion. The standard projection and the
corresponding Lindblad equation are not reliable if higher orders
lead to contributions that are not bounded in time, signifying the
non-uniform convergence of the perturbation expansion \cite{BGM}.

\subsubsection{Conservation laws and relevant
               observables}\label{CONS}
For many physical models one knows certain quantities $C$ which
are exactly (or at least approximately) conserved under the
time-evolution. A great advantage of the formulation by means of
the master equation (\ref{GEN-MASTER}) is that it allows the
implementation of the corresponding conservation laws. For
instance, the master equation derived in Ref.~\cite{GASPARD1}
reflects the conservation of the (uncoupled) total system energy.

To formulate the conservation of $C$ one chooses the operators
$A_i$ in such a way that $C$ is a relevant observable [see
Sec.~\ref{REL-OBS}]. According to Eq.~(\ref{O-REL}) this means
that $C$ can be written in the form
\[
 C=\sum_i C^i_S \otimes A_i.
\]
Then we have the exact relation
\[
 {\mathrm{tr}} \{ C\rho(t) \}
 = {\mathrm{tr}} \{ C {\mathcal{P}} \rho(t) \}
 = \sum_i {\mathrm{tr}}_S \{ C^i_S \rho_i(t) \}.
\]
Hence, we can express the conservation of $C$ on the level of the
master equation by means of the conservation law
\[
 \frac{d}{dt} \sum_i {\mathrm{tr}}_S \{ C^i_S \rho_i(t) \} = 0.
\]
By use of the master equation (\ref{GEN-MASTER}) this yields a
relation between the operators $H^i$ and $R^{ij}_{\lambda}$:
\[
 i \left[ H^i,C^i_S \right] + \sum_{j \lambda}
 \left( R^{ji\dagger}_{\lambda} C^j_S R^{ji}_{\lambda}
 - \frac{1}{2} \left\{ R^{ji\dagger}_{\lambda} R^{ji}_{\lambda},
 C^i_S \right\} \right) = 0.
\]
Thus, known conserved quantities lead to constraints on the choice
of the operators that enter the master equation.

An example is given by the quantity $C=\sigma_+\sigma_- + \Pi_2$
which counts the total number of excitations for the model
discussed in Sec.~\ref{EXAMPLES}. This quantity is exactly
conserved under the evolution generated by the Hamiltonian
(\ref{H-MODEL}). Writing
\[
 C = \sigma_+\sigma_- \otimes \Pi_1 + (\sigma_+\sigma_- + I_S) \otimes \Pi_2
\]
we see that $C$ is indeed a relevant observable for the projection
(\ref{NEWPROJECTION}), i.~e., we have
${\mathcal{P}}^{\dagger}C=C$. The corresponding conservation law
takes the form
\[
 p_e(t) + {\mathrm{tr}}_S \rho_2(t) = {\mathrm{const.}}
\]
This fact can be used to motivate the projection superoperator:
With the choice of Eq.~(\ref{NEWPROJECTION}) one ensures that the
projection superoperator leaves invariant the conserved quantity
and that the effective description respects the corresponding
conservation law.

\section{Conclusions}\label{CONCLU}
We have discussed the theoretical description of non-Markovian
quantum dynamics within the framework of the projection operator
techniques. It has been shown that an efficient modelling of
strong non-Markovian effects is made possible through the use of
correlated projection superoperators ${\mathcal{P}}$ that take
into account statistical correlations between the open system and
its environment. We have formulated and explained some general
physical conditions which demand, essentially, that
${\mathcal{P}}$ can be expressed in terms of a projective quantum
channel that operates on the environmental variables. On the basis
of these conditions a representation theorem for correlated
projection operators [Eq.~(\ref{PROJECTION-GENFORM})] has been
derived.

Employing a correlated projection superoperator instead of the
usual projection onto a tensor product state, one enlarges the set
of dynamical variables from the reduced density matrix $\rho_S$ to
a collection of densities $\rho_i$ describing system states that
are correlated with certain environmental states. By means of an
embedding of the non-Markovian dynamics into a Lindblad dynamics
on a suitably extended state space, we have derived the general
structure of a master equation [Eq.~(\ref{GEN-MASTER})] which
governs the dynamics of the $\rho_i$ and models strong
non-Markovian effects, while preserving the physical conditions of
normalization and positivity. A particularly important feature of
the master equation is that it is able to describe very long and
even infinite memory times, as well as large correlations in the
initial state.

We have also discussed the role of known conserved quantities.
Such quantities may be helpful to find an appropriate projection
superoperator by demanding that they be relevant observables for
${\mathcal{P}}$. Once this has been achieved one can express the
corresponding conservation laws on the level of the effective
description provided by the master equation.

The semigroup assumption used in the derivation of the master
equation (\ref{GEN-MASTER}) is not really necessary. In fact, the
present formulation can easily be extended to the case of an
explicitly time-dependent Lindblad generator on the extended state
space. The resulting master equation is then again of the form of
Eq.~(\ref{GEN-MASTER}), where the operators $H^i(t)$ and
$R_{\lambda}^{ij}(t)$ now depend explicitly on time.

The result expressed by Eq.~(\ref{GEN-MASTER}) could be
particularly useful for a phenomenological approach to
non-Markovian dynamics: For arbitrary $H^i$ and $R^{ij}_{\lambda}$
this equation represents a physically acceptable master equation
because it preserves the normalization of the reduced density
matrix and transforms positive initial components $\rho_i(0) \geq
0$ into positive components $\rho_i(t) \geq 0$. We emphasize that
the argument leading to the form (\ref{GEN-MASTER}) of the master
equations is non-perturbative. Given a certain projection
superoperator ${\mathcal{P}}$ defining the densities $\rho_i$, the
only assumption entering the derivation is the existence of an
embedding of the dynamics into a Lindblad dynamics on the extended
state space.

The projection superoperators used for the description of
non-Markovian dynamics in Sec.~\ref{DYNAMICS} have a special
feature. Namely, they project any given state onto a classically
correlated state. If a non-Markovian dynamics can be approximated
by use of such a superoperator within low orders of the
perturbation expansion, one can conclude that the true states of
the total system can be represented effectively through
classically correlated states and that genuine quantum
correlations (entanglement) may be treated as perturbations. On
the other hand, the representation theorem of
Sec.~\ref{SEC-REPRESENTATION} includes the case of projection
superoperators that project onto nonseparable (entangled) quantum
states. For such superoperators the arguments that led to the
master equation (\ref{GEN-MASTER}) are not applicable. It remains
an important open problem to extend the formulation developed here
to the case of nonseparable projection superoperators. Such an
extension could enable a systematic investigation of the dynamical
significance of genuine quantum correlations in non-Markovian
processes.

\begin{acknowledgments}
I would like to thank Jochen Gemmer, Mathias Michel, and Daniel
Burgarth for stimulating discussions and helpful comments on the
manuscript.
\end{acknowledgments}


\begin{thebibliography}{99}
\bibitem{TheWork} H. P. Breuer and F. Petruccione,
                  \textit{The Theory of Open Quantum Systems}
                  (Oxford University Press, Oxford, 2002).
\bibitem{REIBOLD} F. Haake and R. Reibold, Phys. Rev. A
                  \textbf{32}, 2462 (1985).
\bibitem{HU} B. L. Hu, J. P. Paz and Y. Zhang,
             Phys. Rev. D \textbf{45}, 2843 (1992).
\bibitem{INGOLD} H. Grabert, P. Schramm, and G.-L. Ingold,
                 Phys. Rep. \textbf{168}, 115 (1988).
\bibitem{BUZEK} P. \v{S}telmachovi\v{c} and V. Bu\v{z}ek,
                Phys. Rev. A \textbf{64}, 062106 (2001);
                Phys. Rev. A \textbf{67}, 029902(E) (2003).
\bibitem{LOSS} J. Schliemann, A. Khaetskii, and D. Loss, J. Phys.:
               Condens. Matter \textbf{15}, R1809 (2003).
\bibitem{GEMMER2005} J. Gemmer and M. Michel, Europhys. Lett. \textbf{73}, 1 (2006).
\bibitem{GMM} J. Gemmer, M. Michel, and G. Mahler, \textit{Quantum Thermodynamics},
              Lecture Notes in Physics, Vol. 657 (Springer, Berlin, 2004).
\bibitem{FOURIER} M. Michel, G. Mahler and J. Gemmer, Phys. Rev.
                  Lett. \textbf{95}, 180602 (2005).
\bibitem{NAKAJIMA} S. Nakajima, Progr. Theor. Phys. \textbf{20}, 948 (1958).
\bibitem{ZWANZIG} R. Zwanzig, J. Chem. Phys. \textbf{33}, 1338 (1960).
\bibitem{HAAKE} F. Haake, \textit{Statistical Treatment of Open Systems},
                Springer Tracts in Modern Physics, Vol. 66 (Springer, Berlin, 1973).
\bibitem{KUBO} R. Kubo, M. Toda, and N. Hashitsume,
               \textit{Statistical Physics II. Nonequilibrium Statistical Mechanics},
               (Springer, Berlin, 1991).
\bibitem{SHIBATA} F. Shibata, Y. Takahashi, and N. Hashitsume,
                  J. Stat. Phys. \textbf{17}, 171 (1977);
                  S. Chaturvedi and F. Shibata,
                  Z. Phys. B \textbf{35}, 297 (1979);
                  F. Shibata and T. Arimitsu,
                  J. Phys. Soc. Jap. \textbf{49}, 891 (1980);
                  C. Uchiyama and F. Shibata,
                  Phys. Rev. E \textbf{60}, 2636 (1999);
                  A. Royer, Phys. Lett. A \textbf{315}, 335 (2003);
                  H. P. Breuer, Phys. Rev. A \textbf{70}, 012106 (2004).
\bibitem{GORINI} V. Gorini, A. Kossakowski, and E. C. G. Sudarshan,
                 J. Math. Phys. \textbf{17}, 821 (1976).
\bibitem{LINDBLAD} G. Lindblad, Commun. Math. Phys. \textbf{48}, 119 (1976).
\bibitem{SPOHN} H. Spohn, Rev. Mod. Phys. \textbf{53}, 569 (1980).
\bibitem{BBP} H. P. Breuer, D. Burgarth, and F. Petruccione,
              Phys. Rev. B \textbf{70}, 045323 (2004).
\bibitem{GASPARD1} M. Esposito and P. Gaspard,
                   Phys. Rev. E \textbf{68}, 066112 (2003).
\bibitem{GASPARD2} M. Esposito and P. Gaspard,
                   Phys. Rev. E \textbf{68}, 066113 (2003).
\bibitem{BUDINI05} A. A. Budini, Phys. Rev. E \textbf{72}, 056106 (2005).
\bibitem{BGM} H. P. Breuer, J. Gemmer, and M. Michel,
              Phys. Rev. E \textbf{73}, 016139 (2006).
\bibitem{BARNETT} S. M. Barnett and S. Stenholm,
                  Phys. Rev. A \textbf{64}, 033808 (2001).
\bibitem{BUDINI04} A. A. Budini, Phys. Rev. A \textbf{69}, 042107 (2004).
\bibitem{LIDAR} A. Shabani and D. A. Lidar, Phys. Rev. A \textbf{71}, 020101(R) (2005).
\bibitem{MANISCALCO} S. Maniscalco and F. Petruccione,
                     Phys. Rev. A \textbf{73}, 012111 (2006).
\bibitem{KOSSAKOWSKI} V. Gorini, A. Frigerio, M. Verri, A. Kossakowski, and
                      E. C. G. Sudarshan, Rep. Math. Phys. \textbf{13}, 149 (1978).
\bibitem{KRAUS} K. Kraus, Ann. Phys. (N.Y.) \textbf{64}, 311 (1971).
\bibitem{STINESPRING} W. F. Stinespring,
                      Proc. Amer. Math. Soc. \textbf{6}, 211 (1955).
\bibitem{ALBER} G. Alber, T. Beth, M. Horodecki, P. Horodecki, R.
                Horodecki, M. R\"otteler, H. Weinfurter, R.
                Werner, and A. Zeilinger, \textit{Quantum Information}
                (Springer-Verlag, Berlin, 2001).
\bibitem{WERNER} R. F. Werner, Phys. Rev. A \textbf{40}, 4277 (1989).
\bibitem{BUDINI06} A. A. Budini, Phys. Rev. A \textbf{74}, 053815 (2006).
\end{thebibliography}
\end{document}